# Dynamical properties of a two-level atom in three variants of the two-photon q-deformed Jaynes-Cummings model

**M. H. Naderi, M. Soltanolkotabi and R. Roknizadeh**
*Quantum Optics Group, Department of Physics , University of Isfahan, Isfahan, Iran*

**Abstract**

Temporal evolution of atomic properties including the population inversion and quantum fluctuations of atomic dipole variables are discussed in three various variants of two-photon q-deformed Jaynes-Cummings model. The model is based on the generalized deformed oscillator algebra, $[\hat{A},\hat{A}^+] = (\hat{N}+1)f^2(\hat{N}+1) - \hat{N}f^2(\hat{N})$ in which $f(\hat{N})$ as a function of number operator $\hat{N}$ determines not only the intensity dependence of atom-field coupling when the model Hamiltonian is expressed in terms of nondeformed field operators but also the structure of initial state of the radiation field. With the field initially being in three different types of q-deformed coherent states, each of them corresponding to a particular form of the function $f(\hat{N})$, the quantum collapse and revival effects as well as atomic dipole squeezing are studied for both on-and off-resonant atom-field interaction. Particularly, it is shown that for nonzero detuning the atomic inversion exhibits superstructures, which are, absent in the nondeformed Jaynes-Cummings model and the dipole squeezing may be enhanced

# 1. Introduction

For the past three decades the Jaynes-Cummings model (JCM) [1] has been playing a very significant role in our understanding of the interaction between radiation and matter in quantum optics. This rather simplified but nontrivial model idealizes the real situation by concentrating on the near-resonance linear coupling between a single two-level atomic system and a quantised mode of radiation. Despite being simple enough to be analytically soluble in the rotating-wave approximation, this model has been a long-lasting source of insight into the nuances of the interaction between light and matter. It has led to nontrivial predictions, such as the existence of collapses and revivals in the atomic excitation[2], atomic dipole squeezing[3], vacuum Rabi oscillation[4] and has also allowed a deeper understanding of the dynamical entangling and disentangling of the atom-field system in the course of time[5]. Further interest in the JCM comes from the fact that recent technological advances have enabled us to experimentally realize this rather idealized model[6] and to verify some of the theoretical predictions. A JCM interaction can be experimentally realized in cavity- QED setups[6] and also, as an effective interaction, in laser-cooled trapped ions[7]. For example revivals in the atomic excitation have recently been observed in a cavity-QED experiment [8], providing direct evidence for the discreteness of photons.

Stimulated by the success of the JCM, more and more people have paid special attention to extending and generalizing the model in order to explore new quantum effects[9]. Among the generalized versions, the one with intensity-dependent coupling[10] and the one which describes two-photon (or generally multi-photon) transitions[11] are of special interest. The intensity dependent JCM is interesting since this kind of interaction represents a very simple case of a nonlinear interaction corresponding to a more realistic physical situation and the importance of the two-photon processes is due to the fact that the high degree of correlation between the photons in a pair may lead to the generation of nonclassical states of the radiation field [12] .

On the other hand the quantum algebras[13] introduced as a mathematical description of deformed Lie algebras, depending, in general, on one or more deformation parameters have given the possibility of generalizing the notion of creation and annihilation operators of the usual oscillator and to introduce deformed oscillator. The representation theory of the quantum algebras with a single deformation parameter $q$ has led to the development of the $q$-deformed oscillator algebra[14,15]. Using a q-oscillator description Chaichian and co-workers[16] were the first to generalize the JCM Hamiltonian with an intensity-dependent coupling by relating it to the quantum $su_q(1,1)$ algebra. Similarly, Buzek[17] with the aim of extraction possible information about the physical meaning of the $q$-deformation studied the atomic inversion of the standard JCM with a $q$-deformed field initially prepared in a $q$-deformed coherent state($q$-DCS)[18]. Furthermore, the quantum collapse and revival effects as well as the squeezing properties of the radiation field in the $q$-deformed version of the one-photon on-resonant JCM were investigated by Crnuglej et al [19].

The aim of the present paper is to develop the above-mentioned deformed JCMs and investigate some aspects of three various variants of two-photon $q$-deformed JCM. Formally, the model Hamiltonian for the atom-field system has the same structure as nondeformed two-photon JCM with $\hat{a}$ and $\hat{a}^+$ replaced by the deformed operators $\hat{A}$ and $\hat{A}^+$ obeying the deformed commutation relation $[\hat{A},\hat{A}^+] = (\hat{N}+1)f^2(\hat{N}+1) - \hat{N}f^2(\hat{N})$. The nonlinearity function $f(\hat{N})$ plays a central role in our treatment since it determines the form of nonlinearities of both the field and the intensity-dependent atom-field coupling. With the field initially being in three different types of *q*-deformed coherent states, each of them corresponding to a particular form of the function $f(\hat{N})$, the quantum collapse and revival effects as well as atomic dipole squeezing are studied for both on-and off-resonant atom-field interaction.

Our work is organized as follows. In the next section we describe the theoretical model of multi-photon interaction between a cavity field and a two-level atom within the framework of a generalized deformed JCM with an arbitrary nonlinearity function $f(\hat{N})$ and we construct the eigenvectors and eigenvalues. Section 3 is devoted to give the time- dependent state of the atom- field system initially in an arbitrary state. In section 4, by choosing three different forms of the function $f(\hat{N})$ for which the associated deformed coherent states are known we introduce three various variants of the two-photon q-deformed JCM. The dynamics of atomic properties including the temporal behavior of population inversion and quantum fluctuations of atomic dipole variables are discussed in section 5. Finally, we present a conclusion in section 6.

## 2. The generalized multi-photon deformed JCM

It has been shown [20] that most of the nonlinear generalizations of the JCM are only particular cases of the JCM in which the creation and annihilation operators of the radiation field are replaced by deformed harmonic-oscillator operators with prescribed commutation relations. One possible generalization is the multi-photon deformed JCM ($\hbar = 1$)

$$\hat{H} = \omega \hat{A}^+ \hat{A} + \frac{1}{2}\omega_0 \hat{\sigma}_z + g(\hat{A}^{+m}\hat{\sigma}^- + \hat{A}^m \hat{\sigma}^+) \quad (m=1,2,3,\ldots), \quad (1)$$

Where the two atomic levels separated by an energy difference $\omega_0$ are represented by the Pauli matrices $\hat{\sigma}_z, \hat{\sigma}^\pm$, the coupling constant *g* is a real number and $\omega$ is the frequency of the field. The operators $\hat{A}, \hat{A}^+$ are deformed annihilation and creation operators constructed from the usual bosonic operators $\hat{a}, \hat{a}^+$ ($[\hat{a},\hat{a}^+]=1$) and number operator $\hat{N} = \hat{a}^+ \hat{a}$ as follows

$$\hat{A} = \hat{a}f(\hat{N}) \quad , \quad \hat{A}^+ = f(\hat{N})\hat{a}^+, \quad (2)$$

in which $f(\hat{N})$ is an arbitrary real function of $\hat{N}$. The deformed operators $\hat{A}, \hat{A}^+$ satisfy the deformed bosonic oscillator commutation relations

$$[\hat{A},\hat{A}^+] = \{\hat{N}+1\} - \{\hat{N}\} = (\hat{N}+1)f^2(\hat{N}+1) - \hat{N}f^2(\hat{N}),$$
$$[\hat{A},\hat{N}] = \hat{A} \quad , \quad [\hat{A}^+,\hat{N}] = -\hat{A}^+. \quad (3)$$

The knowledge of the function $f(\hat{N})$ determines all the properties of the deformed algebra (3). It is evident that in the limiting case $f(\hat{N})=1$, the Hamiltonian (1) becomes the multi-photon conventional JC Hamiltonian and the algebra(3) reduces to the well-known Heisenberg-Weyl algebra generated by $\hat{a}, \hat{a}^+$ and the identity $\hat{I}$. Such a generalization is of considerable interest because of its relevance to the study of the intensity-dependent interaction between a single atom and the radiation field with the atom making *m*-photon transitions in quantum optics [4,10,19] as well as the study of the quantized motion of a single ion in an anharmonic-oscillator potential trap[21].

To achieve a more clear insight to the physical meaning of the deformed JC Hamiltonian given in Eq.(1) it is proper to rewrite it in terms of nondeformed field operators $\hat{a}$, $\hat{a}^+$. Using (2) we arrive at

$$\hat{H} = \omega\hat{a}^+\hat{a} + \frac{1}{2}\omega_0\hat{\sigma}_z + \omega\left(f^2(\hat{a}^+\hat{a})-1\right)\hat{a}^+\hat{a} + g\left(\frac{(f(\hat{a}^+\hat{a}))!}{(f(\hat{a}^+\hat{a}-m))!}\hat{a}^{+m}\hat{\sigma}^- + \hat{a}^m\frac{(f(\hat{a}^+\hat{a}))!}{(f(\hat{a}^+\hat{a}-m))!}\hat{\sigma}^+\right)$$
(4)

with $(f(\hat{N}))! = f(\hat{N})f(\hat{N}-1).....f(1)$ and $(f(0))! = 1$. In this way, we learn that the above Hamiltonian describes an intensity–dependent multi-photon coupling between a single two-level atom and a nondeformed single-mode radiation field in the presence of an additional nonlinear interaction represented by the third term in Eq.(4). As a well-known example if we choose

$$f(\hat{a}^+\hat{a}) = \sqrt{1+k(\hat{a}^+\hat{a}-1)},$$
(5)

where $k$ is a positive constant, then the third term in Eq.(4) takes the form $\chi\hat{a}^{+2}\hat{a}^2$ which is reminiscent of the Kerr-induced interaction[22] with $\chi = k\omega$ as the dispersive part of the third-order nonlinearity of the Kerr-like medium. So, in this case the model consists of a single two-level atom undergoing m-photon processes in a single –mode field surrounded by a nonlinear Kerr-like medium contained inside a lossless cavity. Physically, this model may be realized as if the cavity contains two different species of Rydberg atoms, of which one behaves like a two-level atom undergoing m-photon transitions and the other behaves like an anharmonic oscillator in the single mode field of frequency $\omega$[23]. Furthermore it is easily seen that with the choice (5) for the function $f(\hat{N})$ the commutation relations (3) take the form

$$[\hat{A},\hat{A}^+] = 2\hat{A}_0 \quad , \quad [\hat{A}_0,\hat{A}^+] = k\hat{A}^+ \quad , \quad [\hat{A}_0,\hat{A}] = -k\hat{A},$$
(6)

with $\hat{A}_0 = \frac{1}{2} + k\hat{a}^+\hat{a}$. It is noteworthy that the relations (6) define su(1,1) algebra when $k=1$.

The atom field evolution is studied in the space $|e,n\rangle$ and $|g,n\rangle$ where n=0,1,2,…. The state $|e,n\rangle$ ($|g,n\rangle$) means that the atom is in the excited state $|e\rangle$ (ground state $|g\rangle$) and the field in the number state $|n\rangle$ which is the normalized Fock state of the deformed operators $\hat{A}$, $\hat{A}^+$,

$$|n\rangle = \frac{(\hat{A}^+)^n}{\sqrt{\{n\}!}}|0\rangle \quad , \quad \hat{A}|n\rangle = \sqrt{\{n\}}|n-1\rangle \quad , \quad \hat{A}^+|n\rangle = \sqrt{\{n+1\}}|n+1\rangle \quad , \quad \{\hat{N}\}|n\rangle = \hat{A}^+\hat{A}|n\rangle = \{n\}|n\rangle$$
(7)

with $\{n\} = nf^2(n)$. The states $|e,n\rangle$, $|g,n\rangle$ are eigenstates of $\omega\hat{A}^+\hat{A} + \frac{\omega_0}{2}\hat{\sigma}_z$ with the respective eigenvalues $E_{e,n} = \omega\{n\} + \frac{\omega_0}{2}$ and $E_{g,n} = \omega\{n\} - \frac{\omega_0}{2}$. The atom-field interaction part of the Hamiltonian(1) is such that the state $|e,n\rangle$ is taken to $|g,n+m\rangle$ and vice versa, during the evolution of the system. Thus the entire Hilbert space is split into subspaces spanned by $|e,n\rangle$ and $|g,n+m\rangle$ and the dynamics confined to individual subspaces. It is easy to show that the eigenvalues of the Hamiltonian (1) are

$$E_{\pm,n}^{(m)} = \frac{\omega}{2}\left(\{n+m\}+\{n\}\right) \pm \frac{1}{2}\sqrt{\Delta_{n,m}^2 + 4g^2 \frac{\{n+m\}!}{\{n\}!}}, \qquad (8)$$

in which $\Delta = \omega_0 - m\omega$ is the detuning parameter and we have defined

$$\Delta_{n,m} = \Delta - \omega\left(\{n+m\} - \{n\} - m\right). \qquad (9)$$

The corresponding eigenvectors are

$$|+,n\rangle_{(m)} = \cos\vartheta_{n,m}|e,n\rangle + \sin\vartheta_{n,m}|g,n+m\rangle,$$
$$|-,n\rangle_{(m)} = \sin\vartheta_{n,m}|e,n\rangle - \cos\vartheta_{n,m}|g,n+m\rangle, \qquad (10)$$

with

$$\cos\vartheta_{n,m} = \frac{2g\sqrt{\frac{\{n+m\}!}{\{n\}!}}}{\sqrt{(\Omega_{n,m}-\Delta_{n,m})^2 + 4g^2 \frac{\{n+m\}!}{\{n\}!}}},$$

$$\sin\vartheta_{n,m} = \frac{(\Omega_{n,m}-\Delta_{n,m})}{\sqrt{(\Omega_{n,m}-\Delta_{n,m})^2 + 4g^2 \frac{\{n+m\}!}{\{n\}!}}}, \qquad (11)$$

and

$$\Omega_{n,m} = \sqrt{\Delta_{n,m}^2 + 4g^2 \frac{\{n+m\}!}{\{n\}!}} \qquad (12)$$

is the generalized Rabi frequency. For $f(\hat{N})=1$ the function $\Omega_{n,m}$ is reduced to the conventional Rabi frequency in nondeformed JCM. In fact depending on the form of $f(\hat{N})$ this function has a different behavior in comparison with the conventional Rabi frequency that, in turn, it leads to some important consequences for the dynamics of the atom-field system. The energy difference between the levels $E_{+,n}^{(m)}$ and $E_{-,n}^{(m)}$ is

$$\Delta E_n^{(m)} = \Omega_{n,m} = \sqrt{\Delta_{n,m}^2 + 4g^2 \frac{\{n+m\}!}{\{n\}!}}. \qquad (13)$$

The minimum of separation occurs when $\Delta_{n,m} = 0$ and the corresponding difference is $2g\sqrt{\frac{\{n+m\}!}{\{n\}!}}$. It is evident that for $f(\hat{N})=1$ the minimum of energy separation occurs at $\Delta = 0$. Therefore the effect of $f(\hat{N}) \neq 1$ is to shift the value of $\Delta$ at which the minimum separation of eigenenergies occurs.

## 3. The time dependent atom- field state

In order to study the dynamics of the system governed by the generalized deformed JC Hamiltonian (1), first we need to describe the time-dependent state of the system. For this purpose we solve the time dependent Schroedinger equation in the interaction picture

$$i\frac{\partial}{\partial t}|\Psi(t)\rangle = \tilde{H}_{int}|\Psi(t)\rangle, \qquad (14)$$

where $\tilde{H}_{int}$ is the interaction part of the Hamiltonian(1) in the interaction picture easily obtained as

$$\tilde{H}_{int} = \exp\left(i(\omega\hat{A}^+\hat{A} + \frac{\omega_0}{2}\hat{\sigma}_z)t\right)\left(g\hat{A}^{+m}\hat{\sigma}^- + g\hat{A}^m\hat{\sigma}^+\right)\exp\left(-i(\omega\hat{A}^+\hat{A} + \frac{\omega_0}{2}\hat{\sigma}_z)t\right)$$

$$= g\left(\hat{A}^{+m}\hat{\sigma}^- \exp\left(i\left(\omega(\{\hat{N}+m\}-\{\hat{N}\}-m)-\Delta\right)t\right) + \exp\left(-i\left(\omega(\{\hat{N}+m\}-\{\hat{N}\}-m)-\Delta\right)t\right)\hat{A}^m\hat{\sigma}^+\right). \qquad (15)$$

Let us assume that the atom is initially prepared either in the excited state $|e\rangle$ or in the ground state $|g\rangle$,

$$|\psi(0)\rangle_{atom} = \alpha e^{i\phi}|e\rangle + \beta|g\rangle \quad, \quad \alpha^2 + \beta^2 = 1, \qquad (16)$$

where $\phi$ is the relative phase of the two atomic levels, and the field is initially in the superposition of the number states

$$|\psi(0)\rangle_{field} = \sum_{n=0}^{\infty} Q_n |n\rangle. \qquad (17)$$

Therefore the initial state of the atom-field is

$$|\Psi(0)\rangle = |\psi(0)\rangle_{atom} \otimes |\psi(0)\rangle_{field}$$

$$= \sum_{n=0}^{\infty} C_{e,n}(0)|e,n\rangle + C_{g,n}(0)|g,n\rangle, \qquad (18)$$

where

$$C_{e,n}(0) = \alpha e^{i\phi} Q_n \quad, \quad C_{g,n}(0) = \beta Q_n. \qquad (19)$$

At any time $t>0$, the atom- field state is described by the state

$$|\Psi(t)\rangle = \sum_{n=0}^{\infty} C_{e,n}^{(m)}(t)|e,n\rangle + C_{g,n}^{(m)}(t)|g,n\rangle, \qquad (20)$$

in which the coefficients $C_{e,n}^{(m)}(t)$ and $C_{g,n}^{(m)}(t)$ are determined in terms of their initial values given in (19) by solving the Schroedinger equation(14),

$$C_{e,n}^{(m)}(t) = \exp\left(i\frac{\Delta_{n,m}}{2}t\right)\left(\left(\cos(\Omega_{n,m}t/2) - i\frac{\Delta_{n,m}}{\Omega_{n,m}}\sin(\Omega_{n,m}t/2)\right)\alpha\exp(i\phi)Q_n - \frac{2ig}{\Omega_{n,m}}\sqrt{\frac{\{n+m\}!}{\{n\}!}}\sin(\Omega_{n,m}t/2)\beta Q_{n+m}\right),$$

$$C_{g,n}^{(m)}(t) = \exp\left(-i\frac{\Delta_{n-m,m}}{2}t\right)\left(\left(\cos(\Omega_{n-m,m}t/2) + i\frac{\Delta_{n-m,m}}{\Omega_{n-m,m}}\sin(\Omega_{n-m,m}t/2)\right)\beta Q_n \right.$$
$$\left. - \frac{2ig}{\Omega_{n-m,m}}\sqrt{\frac{\{n\}!}{\{n-m\}!}}\sin(\Omega_{n-m,m}t/2)\,\alpha\exp(i\phi)Q_{n-m}\right). \tag{21}$$

The initial distribution of the radiation field is assumed to be the deformed Poisson distribution

$$|Q_n|^2 = \left(\exp_f |z|^2\right)^{-1}\frac{|z|^{2n}}{\{n\}!}, \tag{22}$$

where $z = |z|e^{i\theta}$ and $\exp_f(x) = \sum_{n=0}^{\infty}\frac{x^n}{\{n\}!}$ is the deformed exponential function. In other words it is assumed that the field is initially prepared in a deformed coherent state (DCS) that is the eigenstate of the deformed operator $\hat{A}$ and its characteristics are uniquely determined by the function $f(n)$. Of course the occurrence of DCS rather than the standard one with Poisson distribution appears naturally since our model is deformed. It is important to note that in our treatment, the function $f(n)$ determines not only the algebraic aspects of the deformed JC Hamiltonian (1), but also the structure of the initial DCS of the field.

## 4. Three various variants of q- deformed JCM

In this section we specialize our considerations to the situation where the operator-valued function $f(\hat{N})$ is assumed to be a 4-parameter function as follows

$$f(\hat{N}; p, q, \lambda, \mu) = \sqrt{\frac{1}{\hat{N}}q^{-(\mu+2\lambda(\hat{N}-1))}\frac{1-p^{\hat{N}}}{1-p}}, \tag{23}$$

where the parameters $p, q, \lambda, \mu$ are positive and real numbers. Our above choice for the function $f(\hat{N})$ is such that for some specific values of the parameters we arrive at three different types of q-DCS whose mathematical structures and quantum statistical properties are known:

a) if we put $p = q$, $\mu = \lambda = 0$ then

$$f(\hat{N}) = \sqrt{\frac{1}{\hat{N}}\frac{1-q^{\hat{N}}}{1-q}} \tag{24}$$

and according to (17) and (22) the corresponding DCS is the well-known maths-type q- deformed coherent state suggested first by Arik and Coon [18]

$$|z,q\rangle_{(AC)} = \sum_{n=0}^{\infty} Q_n^{(AC)}|n\rangle, \tag{25}$$

with

$$Q_n^{(AC)} = \frac{1}{\sqrt{\sum_{n=0}^{\infty}\frac{|z|^{2n}}{[n]_q!}}}\frac{z^n}{\sqrt{[n]_q!}} = \left(\exp_q(|z|^2)\right)^{-1/2}\frac{|z|^n e^{in\theta}}{\sqrt{[n]_q!}}, \tag{26}$$

and $[n]_q = (1-q^n)/(1-q)$, $[n]_q! = [n]_q[n-1]_q....[1]_q$. In this case the deformed operators $\hat{A}, \hat{A}^+$ satisfy the algebra $\hat{A}\hat{A}^+ - q\hat{A}^+\hat{A} = 1$. Furthermore the convergence of the series in (26) requires that for $q<1$, $|z|^2 < 1/(1-q)$ and for $q \geq 1$ there is no restriction on $|z|$.

b) If we put $p = 1$, $\mu = 0$, $\lambda = 1$ then

$$f(\hat{N}) = q^{-(\hat{N}-1)} \tag{27}$$

and the corresponding DCS is a new type of q-DCS constructed and studied recently by Penson and Solomon [24]

$$|z,q\rangle_{(PS)} = \sum_{n=0}^{\infty} Q_n^{(PS)} |n\rangle, \tag{28}$$

with

$$Q_n^{(PS)} = \frac{1}{\sqrt{\sum_{n=0}^{\infty} \frac{q^{n(n-1)} |z|^{2n}}{n!}}} \frac{|z|^n e^{in\theta}}{\sqrt{n!}} q^{n(n-1)/2}. \tag{29}$$

The convergence of the infinite series in (29) requires that $0 \leq q \leq 1$ for all values of $z$. In this case the operators $\hat{A}, \hat{A}^+$ satisfy the algebra $\hat{A}\hat{A}^+ - q^{-2}\hat{A}^+\hat{A} = q^{-2\hat{N}}$. It has been shown[24] that for $q<1$ the states $|z,q\rangle_{(PS)}$ are nonclassical that is they exhibit sub-Poissonian photon statistics and quadrature squeezing.

c) Finally, if we put $p = q$, $\mu = 1$, $\lambda = 1/2$ we have

$$f(\hat{N}) = \sqrt{\frac{1}{\hat{N}} \frac{q^{-\hat{N}} - 1}{1-q}} \tag{30}$$

and the corresponding DCS is another new type of q-DCS studied very recently by Quesne[25]

$$|z,q\rangle_{(Q)} = \sum_{n=0}^{\infty} Q_n^{(Q)} |n\rangle, \tag{31}$$

with

$$Q_n^{(Q)} = \frac{1}{\sqrt{\sum_{n=0}^{\infty} \frac{q^{n(n+1)/2} |z|^{2n}}{[n]_q!}}} \frac{|z|^n e^{in\theta}}{\sqrt{[n]_q!}} q^{n(n+1)/4}, \tag{32}$$

in which for $q>1$, $|z|^2 < 1/(q-1)$ and for $q \leq 1$ there is no restriction on $|z|$. In this case the deformed operators $\hat{A}, \hat{A}^+$ satisfy the deformed commutation relation $\hat{A}\hat{A}^+ - q^{-1}\hat{A}^+\hat{A} = q^{-1}$. It has been shown that similar to the states $|z,q\rangle_{(PS)}$ the states $|z,q\rangle_{(Q)}$ exhibit nonclassical properties [25].

For $q=1$ all three q-DCSs (25), (28) and (31) reduce to the conventional coherent state $|z\rangle = e^{-|z|^2/2} \sum_{n=0}^{\infty} \frac{z^n}{\sqrt{n!}} |n\rangle$ with mean number of photons $\bar{n} = |z|^2$.

It is important to note that the single deformation parameter $q$ in one hand determines the structure of initial DCS of the field and on the other hand it may be viewed as a phenomenological constant controlling the strength of intensity – dependent atom-field coupling as well as the interaction between the field and nonlinear medium contained inside the cavity, described by the fourth and third terms in Hamiltonian (4), respectively. As an interesting point we note that when $q = 1 \pm \varepsilon$ $(0 < \varepsilon << 1)$ the functions $f(\hat{N})$ given by (24) and (30) reduce to $f(\hat{N}) \approx \sqrt{1 + \frac{\varepsilon}{2}(\hat{N}-1)}$ and for $q = 1 - \varepsilon$ the function $f(\hat{N})$ given by (27) reduces to $f(\hat{N}) \approx \sqrt{1 + 2\varepsilon(\hat{N}-1)}$ [note that in the case (b), $q \leq 1$]. Therefore one can infer that up to the first order approximation the nonlinearity of the models under consideration may be described as a Kerr-type nonlinearity [see Eq.(5)].

According to Eq. (8), the energy eigenvalues $E_{\pm,n}^{(2)}$ of the two-photon deformed JCM Hamiltonian (1) read as

$$E_{\pm,n}^{(2)} = \frac{\omega}{2}\left((n+2)f^2(n+2) + nf^2(n)\right) \pm$$
$$\frac{1}{2}\sqrt{\left(\Delta - \omega\left((n+2)f^2(n+2) - nf^2(n) - 2\right)\right)^2 + 4g^2(n+1)(n+2)f^2(n+1)f^2(n+2)}$$
(33)

In Figs.1a-e we have shown the variations of $E_{\pm,n}^{(2)}/\omega$ associated with the functions $f(\hat{N})$ given by (24), (27) and (30) for different values of the deformation parameter $q$, with respect to the scaled detuning parameter $\Delta/\omega = (\omega_0/\omega) - 2$. In each figure the dashed and continuous curves represent the energy eigenvalues for $n=1$ and $n=2$, respectively. The diverging eigenvalue separation beyond the minimum separation indicates level repulsion in the eigenvalues of the dressed atom. Furthermore the effect of q-deformation is to shift the value of $\Delta$ where the minimum separation occurs. For $q=1$ the minimum separation occurs at $\Delta = 0$. Of course, as shown in the figures, the magnitude and direction of the displacement of the position of the minimum separation depend on the form of $f(\hat{N})$ and $q<1$ or $q>1$.

## 5. Dynamics of atomic properties
In the present section we are intended to explore the temporal evolution of the atomic properties in three various variants of two-photon q-deformed JCM, presented in the previous section, interacting with three different types of q-DCS described by Eqs.(25),(28) and (31).

### 5.1. Evolution of atomic population inversion
Using the atom-field state $|\Psi(t)\rangle$ given by Eq. (20) the atomic population inversion at time $t$ is obtained as follows

$$\langle \hat{\sigma}_3(t) \rangle = \langle \Psi(t) | \hat{\sigma}_3 | \Psi(t) \rangle = \sum_{n=0}^{\infty} \left( |C_{e,n}^{(2)}(t)|^2 - |C_{g,n}^{(2)}(t)|^2 \right)$$

$$= \alpha^2 \left( 1 + \sum_{n=0}^{\infty} |Q_n|^2 \left( \frac{4g^2 \{n+1\}\{n+2\}}{\Omega_{n,2}^2} (\cos(\Omega_{n,2} t) - 1) \right) \right)$$

$$- \beta^2 \left( 1 + \sum_{n=0}^{\infty} |Q_{n+2}|^2 \left( \frac{4g^2 \{n+1\}\{n+2\}}{\Omega_{n,2}^2} (\cos(\Omega_{n,2} t) - 1) \right) \right)$$

$$- 8g\alpha\beta \sum_{n=0}^{\infty} \frac{|Q_n||Q_{n+2}|}{\Omega_{n,2}} \sqrt{\{n+1\}\{n+2\}} \sin\left(\frac{\Omega_{n,2}}{2} t\right) \left( \sin\left(\frac{\Delta_{n,2}}{2} t + \phi - 2\theta\right) \cos\left(\frac{\Omega_{n,2}}{2} t\right) \right.$$

$$\left. - \frac{\Delta_{n,2}}{\Omega_{n,2}} \cos\left(\frac{\Delta_{n,2}}{2} t + \phi - 2\theta\right) \sin\left(\frac{\Omega_{n,2}}{2} t\right) \right). \tag{34}$$

In Figs.2a-d we have plotted the behavior of $\langle \hat{\sigma}_3(t) \rangle$, as a function of the scaled time $gt$, at the exact resonance ($\Delta = 0$) for an initially excited atomic state ($\alpha = 1, \beta = 0$) and different values of deformation parameter $q$. Fig. 2a shows the time evolution of $\langle \hat{\sigma}_3(t) \rangle$ for nondeformed ($q = 1$) two-photon JCM, and Figs.2b-d correspond to the deformed case ($q \neq 1$) associated with (24), (27) and (30), respectively. As it is seen depending on the value of $q$ one observes different time behavior of $\langle \hat{\sigma}_3(t) \rangle$. In the close vicinity of $q=1$, i.e., $q = 1 \pm \varepsilon$ ($0 < \varepsilon \ll 1$), the atomic inversion exhibits a finite sequence of quantum collapses and revivals, whereas for $q>1$ or $q<1$ these effects are lost and $\langle \hat{\sigma}_3(t) \rangle$ shows chaotic-like behavior, e.g., for $q=1.1$ in Fig.2b (the top curve), $q=0.95, 0.9, 0.85$ in Fig.2c (the second, third and fourth curves from the bottom, respectively) and $q=0.9$ in Fig.2d (the bottom curve). To understand the origin of chaotic-like behavior of $\langle \hat{\sigma}_3(t) \rangle$ we consider the collapse and revival times. The time required for the first collapse and the following revival of the Rabi oscillations, denoted by $t_c$ and $t_r$ respectively, can be estimated approximately. The first revival of the oscillations occurs if at least the terms oscillating with the greatest weights in (34) with $\alpha = 1, \beta = 0$ acquire a phase difference of $2\pi$. Subsequent revivals occur at the phase differences being multiplicities of $2\pi$. Those terms correspond to $n = \bar{n}$ and $n = \bar{n} \pm 1$, where

$$\bar{n} = \sum_{n=0}^{\infty} n |Q_n|^2 = |z|^2 \frac{1}{\exp_f(|z|^2)} \frac{\partial}{\partial |z|^2} \exp_f(|z|^2) \tag{35}$$

is the mean number of photons of the initial field. Thus

$$t_r = \frac{2\pi}{\Omega_{\bar{n}+1,2} - \Omega_{\bar{n},2}} = \frac{2\pi}{\Omega_{\bar{n},2} - \Omega_{\bar{n}-1,2}}. \tag{36}$$

For the inversion to collapse, the oscillations associated with different values of $n$ should be uncorrelated. Since the width of the deformed Poisson distribution $|Q_n|^2$ is where the probability $|Q_n|^2$ is appreciable an estimate of $t_c$ can be obtained from the condition

$$\left( \Omega_{\bar{n}+\delta n, 2} - \Omega_{\bar{n}-\delta n, 2} \right) t_c = 1, \tag{37}$$

where,
$$\delta n = |z| \sqrt{\partial \bar{n} / \partial |z|^2} \tag{38}$$

is the width of initial distribution $|Q_n|^2$. Note that for $q=1$, $\bar{n} = |z|^2$ and $\delta n = \sqrt{\bar{n}} = |z|$ corresponding to the Poisson distribution. The time $t_r$ corresponding to each of the three different initial DCSs (25), (28) and (31) may be obtained as follows:

− maths-type $q$-DCS:

$$t_r^{(AC)} = 2\pi\omega^{-1} q^{-\bar{n}^{(AC)}} (\ln q)^{-1} \frac{\sqrt{\left(q^{\bar{n}^{(AC)}}(1+q)-2\right)^2 + \frac{4g^2}{\omega^2(1-q)^2}\left(1-q^{\bar{n}^{(AC)}+2}\right)\left(1-q^{\bar{n}^{(AC)}+1}\right)}}{\left(q^{\bar{n}^{(AC)}}(1+q)-2\right)(q+1) - \frac{2g^2}{\omega^2(1-q)^2}\left(1+q-2q^{\bar{n}^{(AC)}+2}\right)} \tag{39}$$

where $\bar{n}^{(AC)} = \sum_{n=0}^{\infty} n |Q_n^{(AC)}|^2$.

− Penson-Solomon $q$-DCS:

$$t_r^{(PS)} = 2\pi\omega^{-1} q^{2\bar{n}^{(PS)}} \sqrt{\left(\bar{n}^{(PS)}(q^{-2}-q^2) + 2(q^{-2}-q^{2\bar{n}^{(PS)}})\right)^2 + \frac{4g^2}{\omega^2 q^2}\left(\bar{n}^{(PS)}+1\right)\left(\bar{n}^{(PS)}+2\right)} \times$$

$$\left[\left(\bar{n}^{(PS)}(q^{-2}-q^2) + 2(q^{-2}-q^{2\bar{n}^{(PS)}})\right)\left((q^{-2}-q^2)(1-2\bar{n}^{(PS)}\ln q - 4q^{-2}\ln q)\right) + \frac{2g^2}{\omega^2 q^2}\left(2\bar{n}^{(PS)}+3-4(\bar{n}^{(PS)}+1)(\bar{n}^{(PS)}+2)\ln q\right)\right]^{-1}$$

(40)

where $\bar{n}^{(PS)} = \sum_{n=0}^{\infty} n |Q_n^{(PS)}|^2$.

− Quesne $q$-DCS:

$$t_r^{(Q)} = 2\pi\omega^{-1} q^{(\bar{n}^{(Q)}+2)} (\ln q)^{-1} \frac{\sqrt{\left(q^{-(\bar{n}^{(Q)}+2)}(1+q)-2\right)^2 + \frac{4g^2}{\omega^2(1-q)^2}\left(1+q-2q^{-(\bar{n}^{(Q)}+1)}\right)}}{\left(2-q^{-(\bar{n}^{(Q)}+2)}(1+q)\right)(q+1) + \frac{2g^2}{\omega^2(1-q)^2}\left(1+q-2q^{-(\bar{n}^{(Q)}+1)}\right)} \tag{41}$$

where $\bar{n}^{(Q)} = \sum_{n=0}^{\infty} n |Q_n^{(Q)}|^2$.

Furthermore the collapse time $T_c$ in each case reads as

$$t_C^{(i)} = \frac{t_r^{(i)}}{4\pi\delta n^{(i)}} = \frac{t_r^{(i)}}{4\pi |z|} \frac{1}{\sqrt{\frac{\partial \bar{n}^{(i)}}{\partial |z|^2}}}, \quad (i = AC, PS, Q) \tag{42}$$

Depending on $q<1$ or $q>1$, the revival and collapse times $t_r^{(i)}$, $t_c^{(i)}$ may be increased or decreased over the times expected for $q=1$, i.e., the revival and collapse times of the nondeformed two-photon JCM which are given by

$$t_r^{(JCM)} = \frac{2\pi}{g}\frac{\sqrt{(\bar{n}+1)(\bar{n}+2)}}{(2\bar{n}+3)} \quad \text{and} \quad t_c^{(JCM)} = \frac{t_r^{(JCM)}}{4\pi\sqrt{\bar{n}}}, \text{ with } \bar{n} = |z|^2.$$

As an example, for the values used in the numerical plots of Fig.2 some of the corresponding revival and collapse times are given in table1. As it is seen in each case where the collapse and revival times decrease in comparison with those correspond to the nondeformed case we have $\frac{t_r^{(i)}}{t_r^{(JCM)}} < \frac{t_c^{(i)}}{t_c^{(JCM)}}$ which in turn leads to overlapping subsequent revivals and ultimately occurrence an irregular behavior of time evolution of $\langle\hat{\sigma}_3(t)\rangle$. Furthermore by increasing the deformation the irregular behavior of population inversion becomes much pronounced; not only the amplitude of Rabi oscillations but also the time average of the atomic inversion is affected remarkably.

**Table 1.** The first revival and collapse times in two-photon nondeformed as well as three variants of $q$- deformed JCM.

| $f(n)=1$ | $q=1$ | $t_r^{(JCM)}=31.3803$ | $t_c^{(JCM)}=0.8323$ |
|---|---|---|---|
| $f(n) = \sqrt{(1-q^n)/n(1-q)}$ | $q=1.1$ | $t_r^{(s)}=11.3779$ | $t_c^{(s)}=0.4113$ |
| $f(n) = q^{-(n-1)}$ | $q=0.85$ | $t_r^{(PS)}=1.6504$ | $t_c^{(PS)}=0.1028$ |
|  | $q=0.9$ | $t_r^{(PS)}=2.7803$ | $t_c^{(PS)}=0.1487$ |
|  | $q=0.95$ | $t_r^{(PS)}=8.5441$ | $t_c^{(PS)}=0.3652$ |
| $f(n) = \sqrt{(q^{-n}-1)/n(1-q)}$ | $q=0.9$ | $t_r^{(Q)}=9.7002$ | $t_c^{(Q)}=0.3692$ |

As another important point, it is to be noted that whereas in the absence of deformation ($q=1$) the atomic inversion oscillates around zero (see Fig.2a), in all three deformed models the atomic inversion oscillates around a positive nonzero value (see Figs.2b-d) which means that due to the deformation more energy is stored in the atomic system leading to energy inhibition. In other words by increasing the deformation there is a growing tendency of the atom to trap the excitation energy. Physically it is due to the change in energy-level structure of the deformed models under consideration.

We now examine the influence of the detuning $\Delta$ on the time evolution of $\langle\hat{\sigma}_3(t)\rangle$. As mentioned in Sec.2 the generalized Rabi frequency $\Omega_{n,m}$ [Eq.(12)] has a different behavior compared to the nondeformed one. In fact, unlike the nondeformed Rabi frequency, the dependence of $\Omega_{n,m}$ on $n$ is such that it has a minimum value. For the two-photon case the function $\Omega_{n,2}$ attains its minimum when detuning satisfies the following relation with $n$

$$\Delta = \omega(\{n+2\} - \{n\} - 2) + \frac{2g^2}{\omega} \frac{\left(\{n+2\}\frac{d}{dn}\{n+1\} + \{n+1\}\frac{d}{dn}\{n+2\}\right)}{\left(\frac{d}{dn}\{n+2\} - \frac{d}{dn}\{n\}\right)},$$
(43)

provided $f(n) \neq 1$ for each value of $n \geq 0$. Choosing $n = \bar{n}$ (mean number of photons in the initial $q$-DCS) we have

$$\Delta = \Delta_c = \omega(\{\bar{n}+2\} - \{\bar{n}\} - 2) + \frac{2g^2}{\omega} \frac{\left(\{\bar{n}+2\}\frac{d}{dn}\{n+1\}\bigg|_{n=\bar{n}} + \{\bar{n}+1\}\frac{d}{dn}\{n+2\}\bigg|_{n=\bar{n}}\right)}{\left(\frac{d}{dn}\{n+2\}\bigg|_{n=\bar{n}} - \frac{d}{dn}\{n\}\bigg|_{n=\bar{n}}\right)}.$$
(44)

The Rabi frequency $\Omega_{n,2}$ in such a case reads

$$\Omega_{n,2} = \sqrt{(\Delta_c - \omega(\{n+2\} - \{n\} - 2))^2 + 4g^2\{n+1\}\{n+2\}}.$$
(45)

Let us treat this frequency as a continuous quantity and express the dispersion curve $\Omega_{n,2}$ around point $\bar{n}$

$$\Omega_{n,2} = \Omega_{\bar{n},2} + (n-\bar{n})^2 \Omega^{(2)}_{\bar{n},2} + \ldots\ldots,$$
(46)

where

$$\Omega^{(2)}_{\bar{n},2} = \frac{1}{2!}\frac{d^2}{dn^2}\Omega_{n,2}\bigg|_{n=\bar{n}} = \frac{1}{2}\frac{\left(\frac{dA(n)}{dn}\right)^2_{n=\bar{n}} - \frac{2g^2}{\omega}\left(B(n)\frac{d^2A(n)}{dn^2} - \frac{d^2C(n)}{dn^2}\right)_{n=\bar{n}}}{\Omega_{\bar{n},2}},$$
(47)

with

$$A(n) = \{n+2\} - \{n\}, \quad B(n) = \frac{\{n+2\}\frac{d}{dn}\{n+1\}\bigg|_{n=\bar{n}} + \{n+1\}\frac{d}{dn}\{n+2\}\bigg|_{n=\bar{n}}}{\frac{d}{dn}\{n+2\}\bigg|_{n=\bar{n}} - \frac{d}{dn}\{n\}\bigg|_{n=\bar{n}}},$$

$$C(n) = \frac{\{n+1\}\{n+2\}}{(n+1)(n+1)} = f^2(n+1)f^2(n+2).$$
(48)

The first term of the expansion (46) is responsible for rapid oscillations of the model under consideration while the remaining terms are responsible for their envelope. If one can neglect the higher order derivatives than the second order one in (46) the collapses and revivals of the oscillations will be perfectly periodic. In other words, by making an appropriate choice of parameters $g$, $q$ and $\bar{n}$ the approximate values predicted by the expression (46) up to second order may match with those given by the exact expression (45) and in such a case it is reasonable to expect regular oscillations with a neat envelope. Under the circumstances, one deals with the second–order revivals [26] which are different from the revivals exhibited by the standard JCM or by the deformed JCM if the condition (44) is not satisfied. Some interesting aspects of the second-order revivals at an initially strongly squeezed radiation field in the standard JCM have recently been considered [27].

In Figs.3a-e we have displayed the effect of detuning on the behavior of $\langle \hat{\sigma}_3(t) \rangle$ for an initially excited atom in each of the three variants of two-photon q-deformed JCM. For each case the values of the parameters are so chosen that the difference between the exact expression (45) and the expansion (46) up to second- order becomes small as far as possible when $\Delta \neq 0$. For an initially maths-type $q$-DCS with $q$=0.9 and $q$=1.1 (Figs. 3a and 3b, respectively) the envelope of $\langle \hat{\sigma}_3(t) \rangle$ when detuning equals to the corresponding $\Delta_c$ is distinct with structures repeating without much distortion, whereas the resonant case reveals irregularity in the time evolution of population inversion. This irregularity is due to the noticeable difference between (45) and (46) when $\Delta = 0$ such that higher order derivatives in (46) lead to incompleteness and overlapping of the revivals. The same situation occurs for an initially Quesne $q$- DCS (Figs.3c and 3d). But for an initially Penson –Solomon $q$-DCS (Fig.3e) the time evolution of $\langle \hat{\sigma}_3(t) \rangle$ exhibits random oscillations when $\Delta = \Delta_c$ and does not have a neat envelope (the top curve in Fig. 3e). Therefore, unlike the two former cases in this case the second-order revivals do not exist and instead of them, regular structures occur at $\Delta \neq \Delta_c \neq 0$ (The middle curve in Fig.3e). In any case, the possibility of regular dynamics of the off-resonant model appears when the resonant model reveals irregularity in its time evolution

## 5.2. Atomic dipole squeezing

To analyze the quantum fluctuations of atomic dipole variables and examine their squeezing we define the two slowly varying Hermitian quadrature operators

$$\hat{\sigma}_1 = \frac{1}{2}\left(\hat{\sigma}^+ e^{-i\omega_0 t} + \hat{\sigma}^- e^{i\omega_0 t}\right), \tag{49}$$
$$\hat{\sigma}_2 = \frac{1}{2i}\left(\hat{\sigma}^+ e^{-i\omega_0 t} - \hat{\sigma}^- e^{i\omega_0 t}\right).$$

In fact $\hat{\sigma}_1$ and $\hat{\sigma}_2$ correspond to the dispersive and absorptive components of the amplitude of the atomic polarization [2], respectively. They obey the commutation relation $[\hat{\sigma}_1, \hat{\sigma}_2] = i\hat{\sigma}_3/2$. Correspondingly, the Heisenberg uncertainty relation is

$$(\Delta \hat{\sigma}_1)^2 (\Delta \hat{\sigma}_2)^2 \geq \frac{1}{16}|\langle \hat{\sigma}_3 \rangle|^2, \tag{50}$$

where $(\Delta \hat{\sigma}_i)^2 = \langle \hat{\sigma}_i^2 \rangle - \langle \hat{\sigma}_i \rangle^2$ is the variance in the component $\hat{\sigma}_i$ ($i$=1, 2) of the atomic dipole.

The fluctuations in the component $\hat{\sigma}_i$ ($i$=1 or 2) are said to be squeezed (i.e., dipole squeezing) if the variance in $\hat{\sigma}_i$ satisfies the condition

$$(\Delta \hat{\sigma}_i)^2 < \frac{1}{4}|\langle \hat{\sigma}_3 \rangle|, \quad (i = 1 \, or \, 2). \tag{51}$$

Since $\langle \hat{\sigma}_i^2 \rangle = 1/4$ this condition may be written as

$$F_i = 1 - 4\langle \hat{\sigma}_i \rangle^2 - |\langle \hat{\sigma}_3 \rangle| < 0, \quad (i = 1 \, or \, 2). \tag{52}$$

The expectation values of the atomic operators $\hat{\sigma}_{1,2}$ in the state $|\Psi(t)\rangle$ of the atom-field system [Eq. (20)] are given by

$$\langle \hat{\sigma}_1 \rangle = \frac{1}{2} \sum_{n=0}^{\infty} \left( C_{e,n}^{(2)*}(t) C_{g,n}^{(2)}(t) e^{-i\omega_0 t} + C_{g,n}^{(2)*}(t) C_{e,n}^{(2)}(t) e^{i\omega_0 t} \right)$$

$$= \sum_{n=0}^{\infty} \alpha^2 U_n(t,\theta) + \beta^2 V_n(t,\theta) + \alpha\beta W_n(t,\phi,\theta),$$

$$\langle \hat{\sigma}_2 \rangle = \frac{1}{2i} \sum_{n=0}^{\infty} \left( C_{e,n}^{(2)*}(t) C_{g,n}^{(2)}(t) e^{-i\omega_0 t} - C_{g,n}^{(2)*}(t) C_{e,n}^{(2)}(t) e^{i\omega_0 t} \right)$$

$$= \sum_{n=0}^{\infty} \alpha^2 U_n(t,\theta+\pi/4) + \beta^2 V_n(t,\theta+\pi/4) + \alpha\beta W_n(t,\phi+\pi/2,\theta),$$

(53)

where
$U_n(t,\theta) =$

$$2g|Q_n||Q_{n+2}| \frac{\sqrt{\{n+1\}\{n+2\}}}{\Omega_{n,2}} \sin\left(\frac{\Omega_{n,2}t}{2}\right) \left( \frac{\Delta_{n+2,2}}{\Omega_{n+2,2}} \sin\left(\frac{\Omega_{n+2,2}t}{2}\right) \cos\left(\frac{(\Delta_{n+2,2}+\Delta_{n,2}+2\omega_0)t}{2} + 2\theta\right) \right.$$

$$\left. - \sin\left(\frac{(\Delta_{n+2,2}+\Delta_{n,2}+2\omega_0)t}{2} + 2\theta\right) \cos\left(\frac{\Omega_{n+2,2}t}{2}\right) \right),$$

$V_n(t,\theta) =$

$$2g|Q_n||Q_{n+2}| \frac{\sqrt{\{n+1\}\{n+2\}}}{\Omega_{n,2}} \sin\left(\frac{\Omega_{n,2}t}{2}\right) \left( \cos\left(\frac{\Omega_{n-2,2}t}{2}\right) \sin\left(\frac{(\Delta_{n,2}+\Delta_{n-2,2}+2\omega_0)t}{2} + 2\theta\right) \right.$$

$$\left. - \frac{\Delta_{n-2,2}}{\Omega_{n-2,2}} \cos\left(\frac{(\Delta_{n,2}+\Delta_{n-2,2}+2\omega_0)t}{2} + 2\theta\right) \sin\left(\frac{\Omega_{n-2,2}t}{2}\right) \right),$$

$$W_n(t,\phi,\theta) = |Q_n|^2 \left( \cos\left(\frac{(\Delta_{n,2}+\Delta_{n-2,2}+2\omega_0)t}{2} + \phi\right) \left( \cos\left(\frac{\Omega_{n,2}}{2}t\right) \cos\left(\frac{\Omega_{n-2,2}}{2}t\right) - \right. \right.$$

$$\frac{\Delta_{n,2}\Delta_{n-2,2}}{\Omega_{n,2}\Omega_{n-2,2}} \sin\left(\frac{\Omega_{n,2}}{2}t\right) \sin\left(\frac{\Omega_{n-2,2}}{2}t\right) \right) + \sin\left(\frac{(\Delta_{n,2}+\Delta_{n-2,2}+2\omega_0)t}{2} + \phi\right) \left( \frac{\Delta_{n-2,2}}{\Omega_{n-2,2}} \cos\left(\frac{\Omega_{n,2}}{2}t\right) \sin\left(\frac{\Omega_{n-2,2}}{2}t\right) + \right.$$

$$\left. \frac{\Delta_{n,2}}{\Omega_{n,2}} \sin\left(\frac{\Omega_{n,2}}{2}t\right) \cos\left(\frac{\Omega_{n-2,2}}{2}t\right) \right) \right) +$$

$$4g^2 \frac{\sqrt{\{n+1\}\{n+2\}\{n+3\}\{n+4\}}}{\Omega_{n,2}\Omega_{n+2,2}} |Q_n||Q_{n+4}| \sin\left(\frac{\Omega_{n,2}}{2}t\right) \sin\left(\frac{\Omega_{n+2,2}}{2}t\right) \cos\left(\frac{(\Delta_{n,2}+\Delta_{n+2,2}+2\omega_0)t}{2} - \phi + 4\theta\right).$$

(54)

The short-time evolution of $F_1(t)$ corresponding to the squeezing of $\hat{\sigma}_1$ have been shown in Figs.4a-c for an initially prepared de-excited atomic state ($\alpha = 0, \beta = 1$) interacting resonantly with initially three types of $q$-DCS (25), (28) and (31) with $|z|^2 = 0.5$ (small mean number of photons) within the framework of three corresponding variants of deformed JCM, respectively. For comparison, in each figure we have also displayed the numerical plot associated with nondeformed case $q=1$. As it is seen, in all three deformed cases the strength of dipole squeezing is reduced compared to that of the nondeformed case. With increasing the deformation the dipole squeezing is weakened considerably.

In Figs.5a-c we have shown the influence of detuning on the time evolution of the function $F_1(t)$. In each figure, the bottom and top curves correspond to the deformed and nondeformed cases respectively, and the value of detuning is chosen such that the corresponding Rabi frequency attains its minimum ($\Delta = \Delta_c$). We find that whereas in the nondeformed case large detuning causes to disappear dipole squeezing completely, for two deformed cases corresponding to the functions $f(n)$ given by (24) and (30) it leads to enhanced and strong dipole squeezing (Figs.5a,5b). The situation is completely changed when we consider $f(n)$ given by (27), the function $F_1(t)$ shows rapid oscillations in an irregular manner and dipole squeezing does not appear at all (bottom curve in Fig.5c).

## 6. Conclusions

In this paper we have studied the nondissipative dynamics of a two-level atom interacting with cavity field through multi-photon transitions within the framework of the deformed JCM corresponds to the JCM including general form of nonlinearity of both the field and the intensity-dependent atom-field coupling. The model is based on the generalized deformed oscillator algebra $[\hat{A}, \hat{A}^+] = (\hat{N}+1)f^2(\hat{N}+1) - \hat{N}f^2(\hat{N})$ in which the function $f(\hat{N})$ determines the nonlinearity of the model under consideration. With the field initially being in three different types of $q$-DCS, each of them corresponding to a particular form of the function $f(\hat{N})$, the time evolution of the atomic inversion and atomic dipole squeezing of three various variants of two-photon $q$-deformed JCM are discussed for both on- and off- resonant atom-field coupling. Particularly, we find that at the exact resonance the atomic inversion exhibits a chaotic-like behavior and the strength of dipole squeezing is reduced compared to the nondeformed case. While for nonzero detuning the dynamical behavior of atomic inversion exhibits superstructures which are absent in the usual JCM and the dipole squeezing may be enhanced.

**Figure Captions**

**Fig.1** Dependence of scaled energy eigenvalues $E_\pm/\omega$ on scaled detuning $\Delta/\omega$ for three various variants of two-photon $q$-deformed JCM with $g/\omega = .5$. The dashed and continuous curves correspond to $n=1$ and $n=2$, respectively.

a, b) $f(n) = \sqrt{\dfrac{1-q^n}{n(1-q)}}$ with $q=0.8$, $q=1.2$ respectively,

c) $f(n) = q^{-(n-1)}$ with $q=0.8$,

d, e) $f(n) = \sqrt{\dfrac{q^{-n}-1}{n(1-q)}}$ with $q=0.8$, $q=1.2$, respectively.

**Fig.2** Time dependence of population inversion for a two-level atom undergoing two-photon transitions. We have set $\alpha^2 = 1$, $g/\omega = 0.1$, $|z|^2 = 9$, $\Delta = 0$.

a) $f(n)=1$,

b) $f(n) = \sqrt{\dfrac{1-q^n}{n(1-q)}}$. The curves from bottom to top of the figure show respectively, $\langle \hat{\sigma}_3(t) \rangle$ for $q=0.9$, $\langle \hat{\sigma}_3(t) \rangle +1.5$ for $q=0.99$, $\langle \hat{\sigma}_3(t) \rangle +3$ for $q=1.01$ and $\langle \hat{\sigma}_3(t) \rangle +4$ for $q=1.1$,

c) $f(n) = q^{-(n-1)}$. The curves from bottom to top of the figure show respectively, $\langle \hat{\sigma}_3(t) \rangle$ for $q=0.99$, $\langle \hat{\sigma}_3(t) \rangle +0.1$ for $q=0.95$, $\langle \hat{\sigma}_3(t) \rangle +0.15$ for $q=0.9$ and $\langle \hat{\sigma}_3(t) \rangle +0.25$ for $q=0.85$,

d) $f(n) = \sqrt{\dfrac{q^{-n}-1}{n(1-q)}}$. The curves from bottom to top of the figure show respectively, $\langle \hat{\sigma}_3(t) \rangle$ for $q=0.9$, $\langle \hat{\sigma}_3(t) \rangle +1.5$ for $q=0.99$, $\langle \hat{\sigma}_3(t) \rangle +3$ for $q=1.01$ and $\langle \hat{\sigma}_3(t) \rangle +4$ for $q=1.1$.

**Fig.3** The influence of detuning on time- evolution of atomic population inversion. Values of $g/\omega$ and $\alpha^2$ are the same as fig.2.

a) $f(n) = \sqrt{\dfrac{1-q^n}{n(1-q)}}$ with $q=0.9$, $|z|^2=5$ ($\bar{n}=7.0963$). The bottom curve shows $\langle \hat{\sigma}_3(t) \rangle -0.5$ for $\Delta = 0$ and the top curve shows $\langle \hat{\sigma}_3(t) \rangle +0.5$ for $\Delta/\omega = \Delta_c/\omega = -2.1737$.

b) $f(n) = \sqrt{\dfrac{1-q^n}{n(1-q)}}$ with $q=1.1$, $|z|^2=2.5$ ($\bar{n}=2.2425$). The bottom curve shows $\langle \hat{\sigma}_3(t) \rangle -0.5$ for $\Delta = 0$ and the top curve shows $\langle \hat{\sigma}_3(t) \rangle +0.5$ for $\Delta/\omega = \Delta_c/\omega = 1.53978$.

c) $f(n) = \sqrt{\dfrac{q^{-n}-1}{n(1-q)}}$ with $q=1.1$, $|z|^2=4$ ($\bar{n}=5.70176$). The bottom curve shows $\langle \hat{\sigma}_3(t) \rangle -0.5$ for $\Delta = 0$ and the top curve shows $\langle \hat{\sigma}_3(t) \rangle +0.5$ for $\Delta/\omega = \Delta_c/\omega = -1.98646$.

d) $f(n) = \sqrt{\dfrac{q^{-n}-1}{n(1-q)}}$ with $q=0.9$, $|z|^2=3$ ($\bar{n}=2.3763$). The bottom curve shows $\langle \hat{\sigma}_3(t) \rangle -0.5$ for $\Delta = 0$ and the top curve shows $\langle \hat{\sigma}_3(t) \rangle +0.5$ for $\Delta/\omega = \Delta_c/\omega = 2.0177$.

e) $f(n) = q^{-(n-1)}$ with $q=0.95$, $x=2.5$ ($\bar{n}=2.0443$). The curves from bottom to top of the figure show respectively, $\langle \hat{\sigma}_3(t) \rangle$ for $\Delta=0$, $\langle \hat{\sigma}_3(t) \rangle$ for $\Delta/\omega = 2$, $\langle \hat{\sigma}_3(t) \rangle +0.2$ for $\Delta/\omega = \Delta_c/\omega = -1.4805$.

**Fig.4** Time evolution of $F_1(t)$ corresponding to the squeezing of $\hat{\sigma}_1$. We have set $\alpha^2 = 0$, $g/\omega = 0.1$, $|z|^2 = 0.5$, $\Delta = 0$.

a) $f(n) = \sqrt{\dfrac{1-q^n}{n(1-q)}}$ with $q= 0.8$ (heavy line), $q=1.2$ (full line).

b) $f(n) = q^{-(n-1)}$ with $q= 0.9$ (heavy line), $q=0.8$ (full line).

c) $f(n) = \sqrt{\dfrac{q^{-n}-1}{n(1-q)}}$ with $q= 0.8$ (heavy line), $q=1.2$ (full line).

In each of these three figures, the dashed curve corresponds to the nondeformed case, $f(n)=1$ ($q=1$).

**Fig.5** Time evolution of the function $F_1(t)$ for $\alpha^2 = 0$, $g/\omega = 0.1$ and nonzero detuning. In each part, the bottom and top curves correspond to the deformed case ($q \neq 1$) and nondeformed case ($q=1$), respectively with the same values of $|z|^2$ and detuning parameter.

a) $f(n) = \sqrt{\dfrac{1-q^n}{n(1-q)}}$ with $q=0.95$, $|z|^2=18$ ($\bar{n} = 49.767$), $\Delta/\omega = \Delta_c/\omega = -8.9003$.

b) $f(n) = \sqrt{\dfrac{q^{-n}-1}{n(1-q)}}$ with $q=1.05$, $|z|^2=18$ ($\bar{n} = 51.650$), $\Delta/\omega = \Delta_c/\omega = -9.2523$.

c) $f(n) = q^{-(n-1)}$ with q=0.95, $|z|^2=18$ ($\bar{n} = 8.0588$), $\Delta/\omega = \Delta_c/\omega = -1.0372$

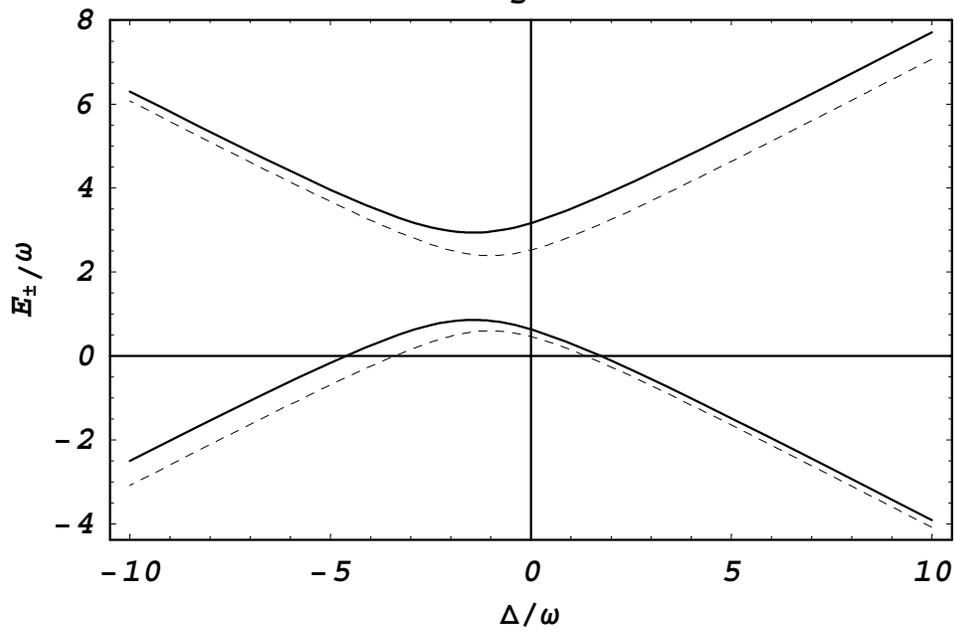

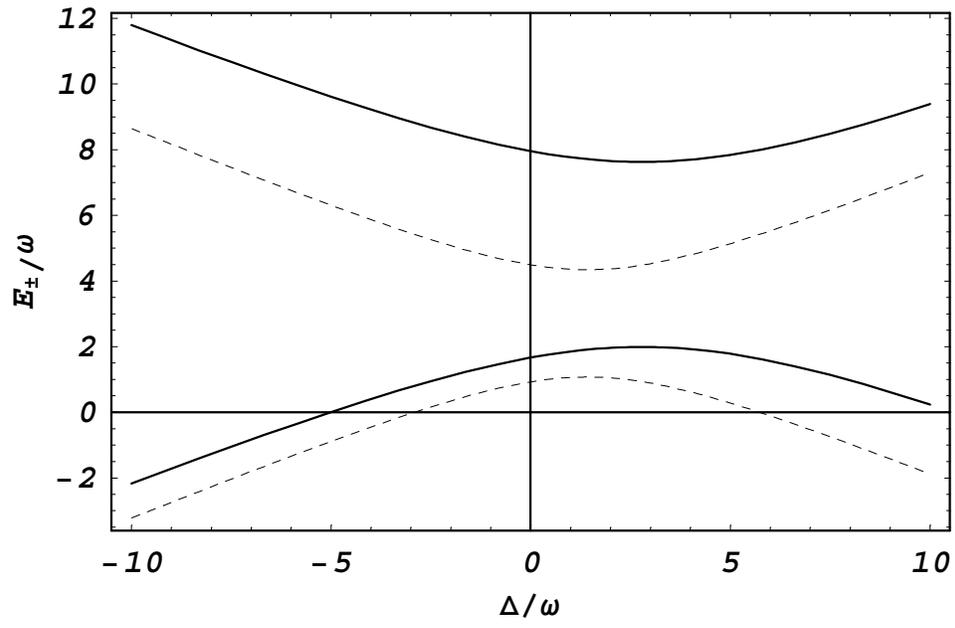

Fig.1b

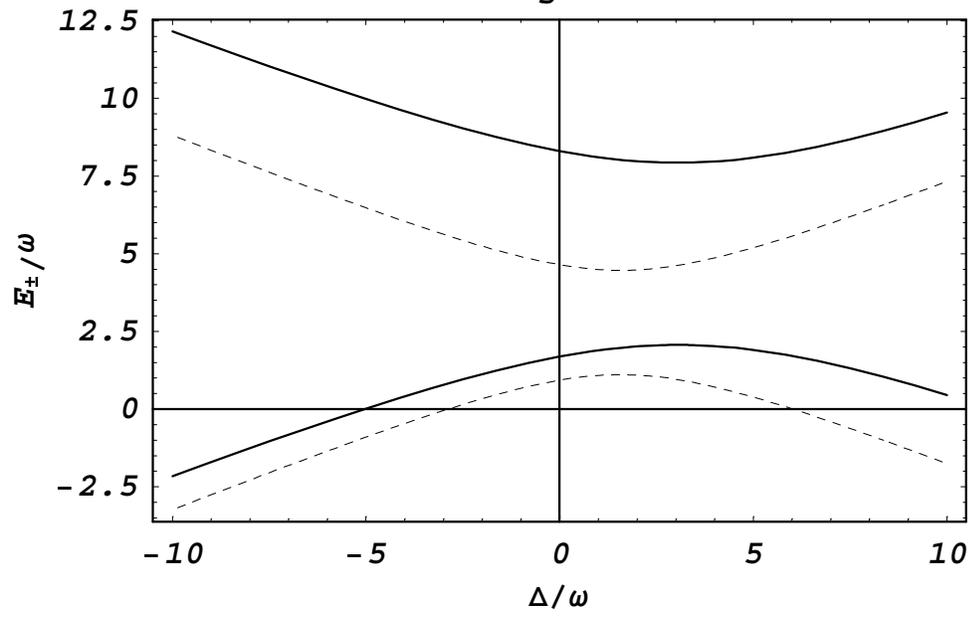

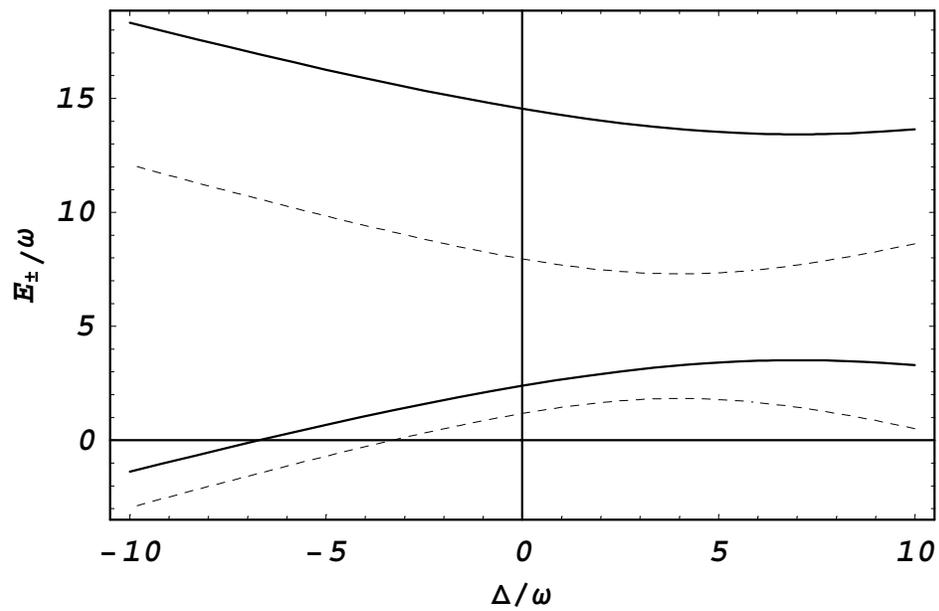

Fig.1d

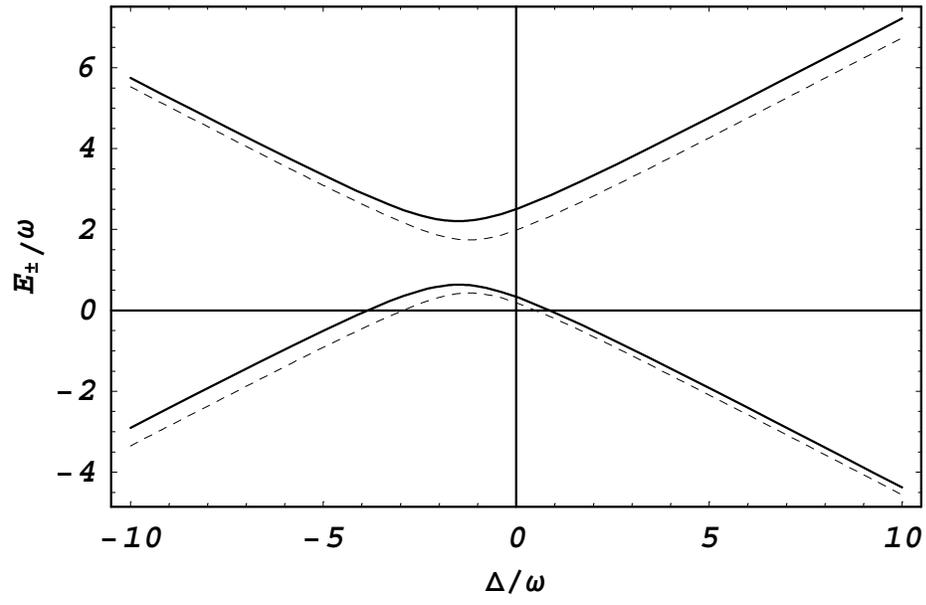

Fig.1e

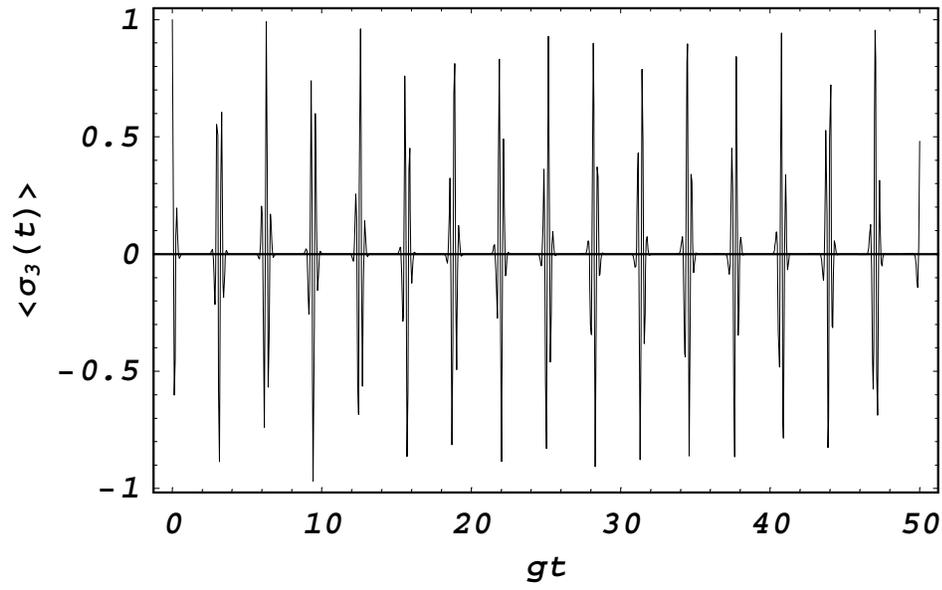

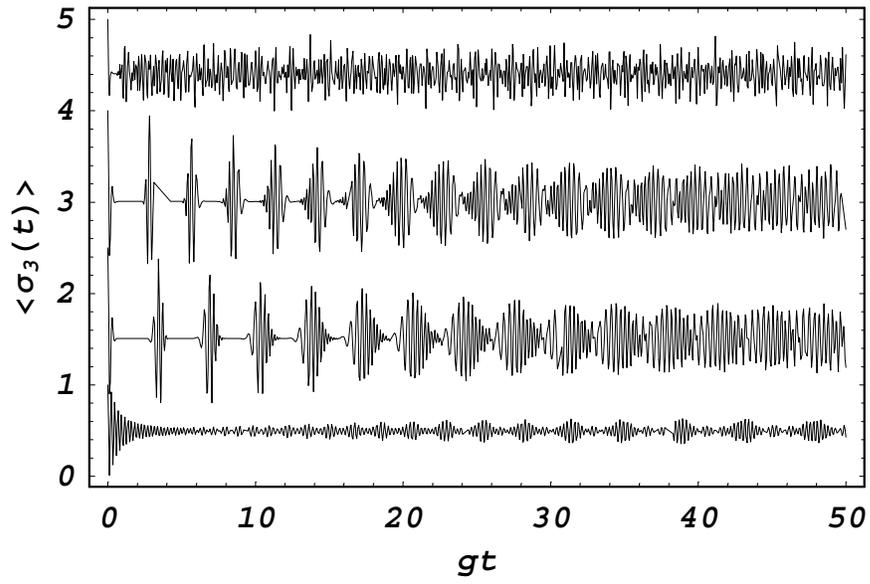

Fig.2b

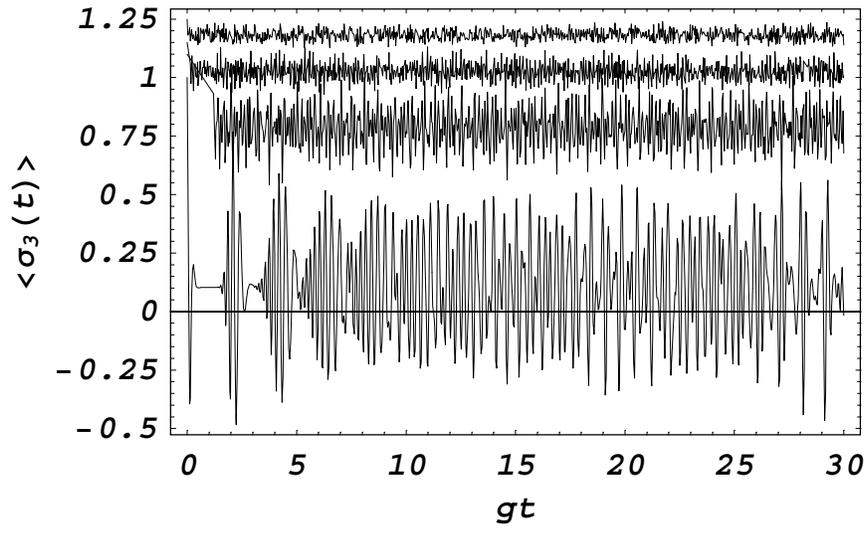

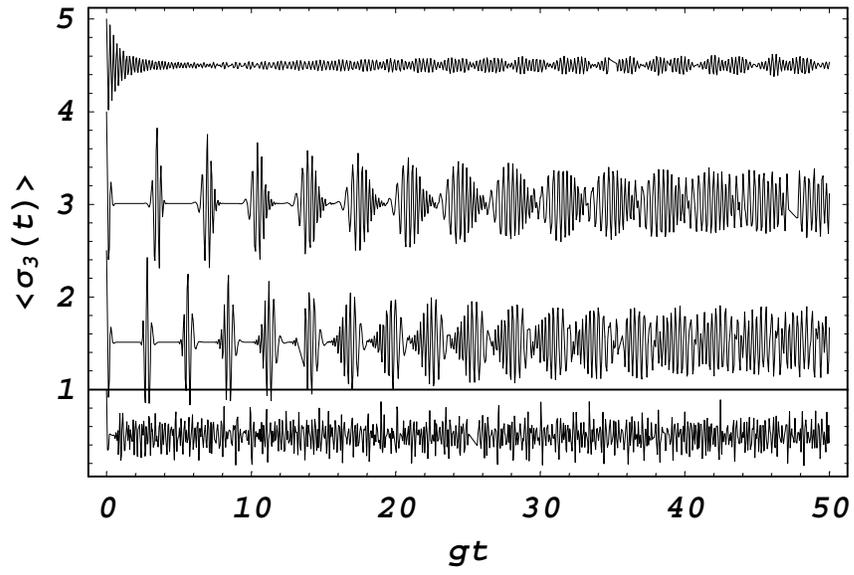

Fig.2d

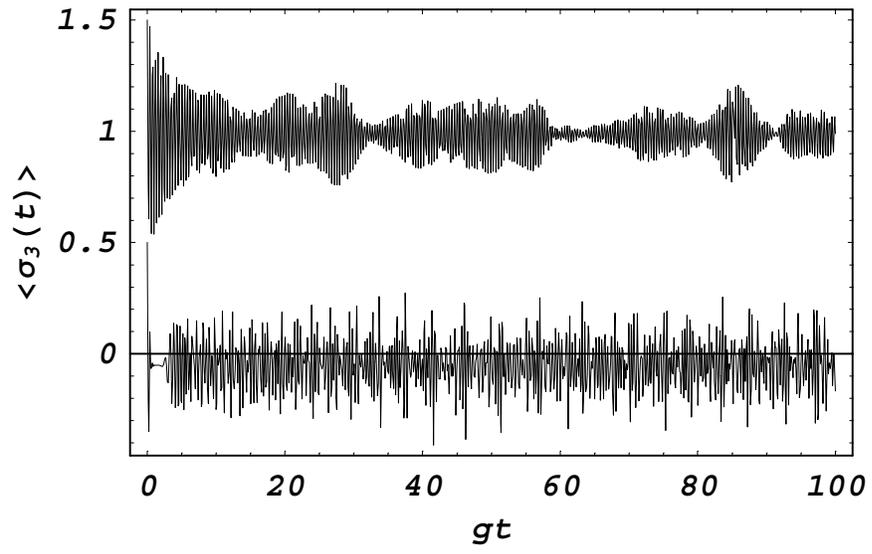

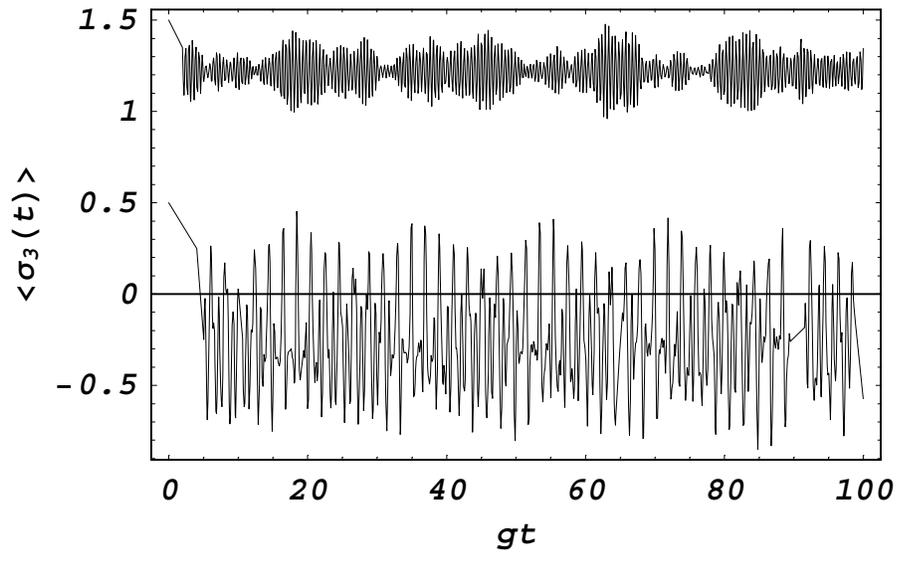

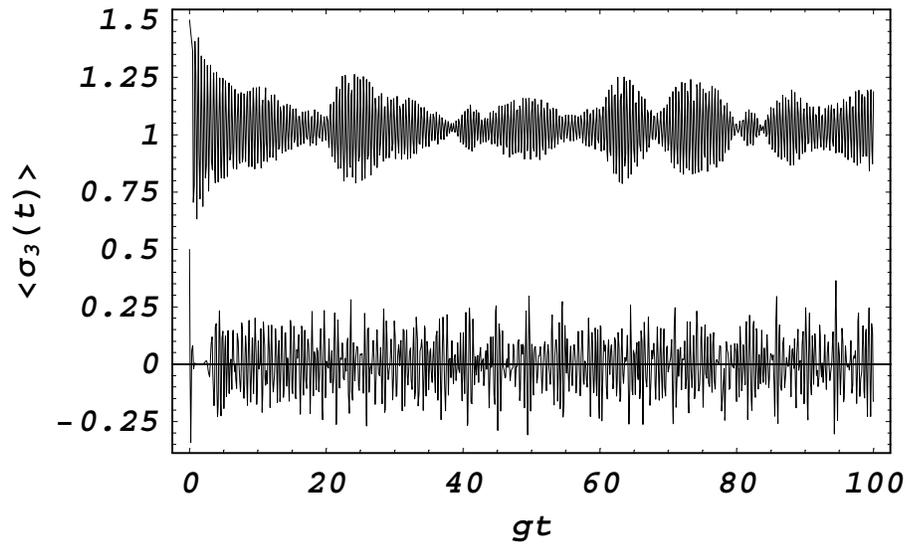

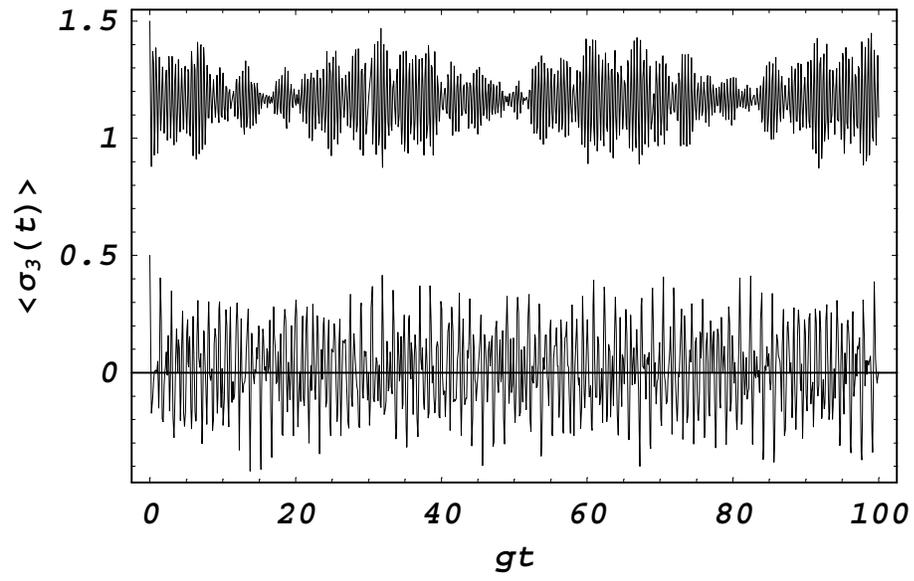

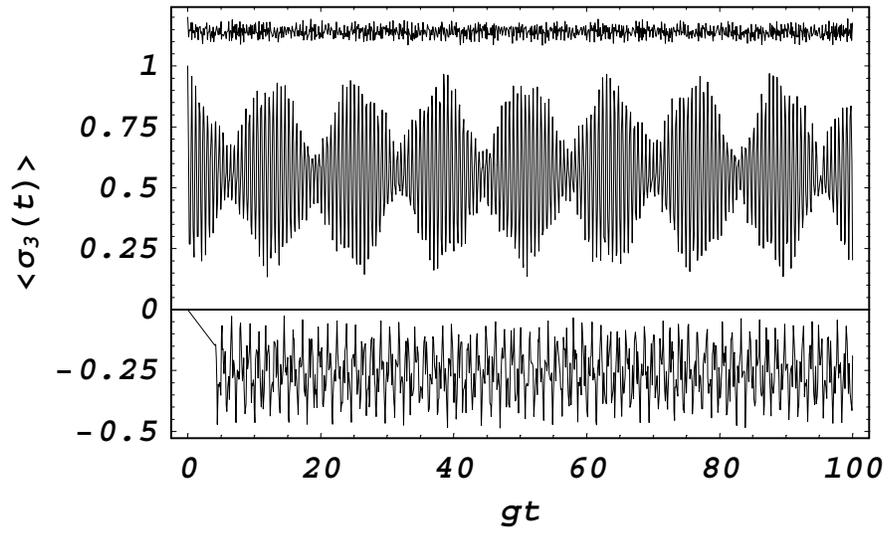

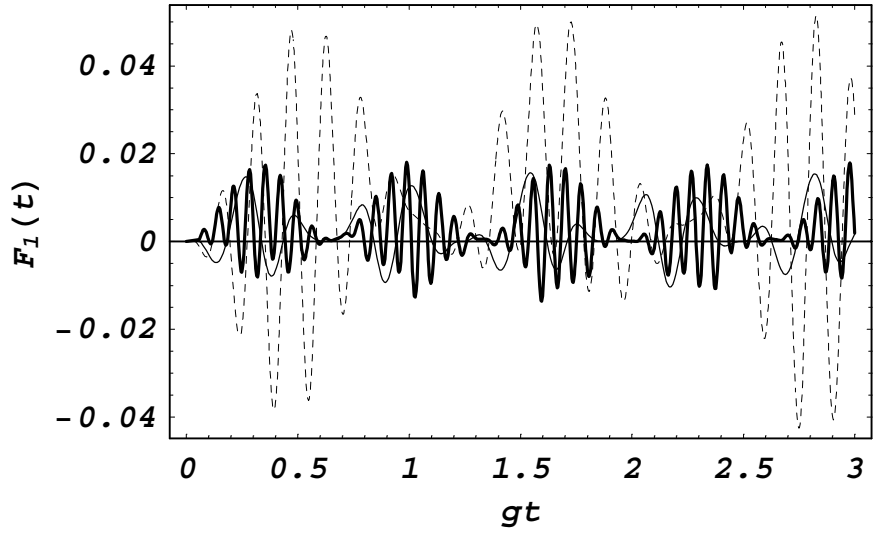

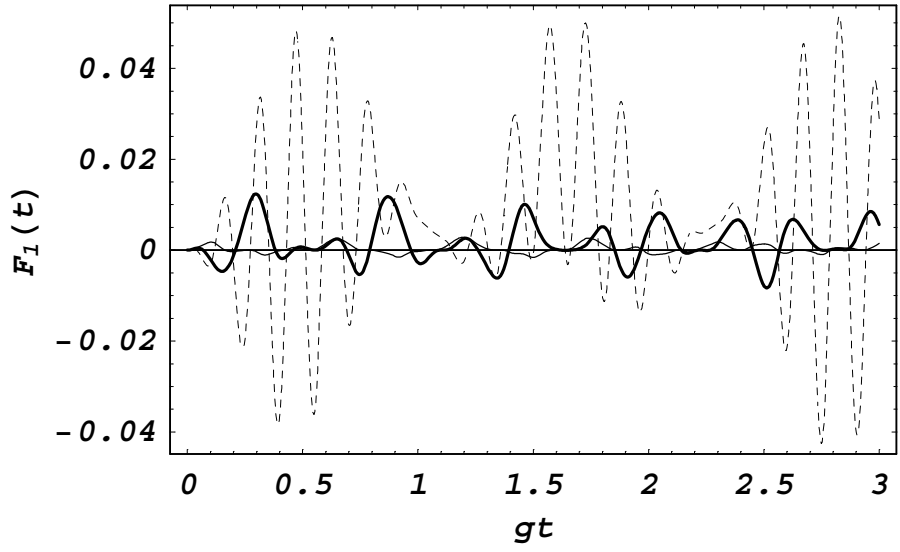

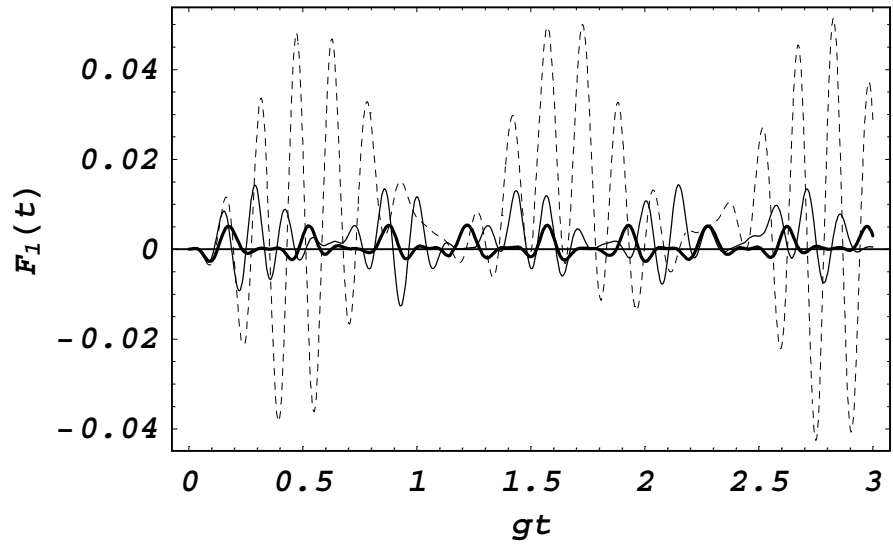

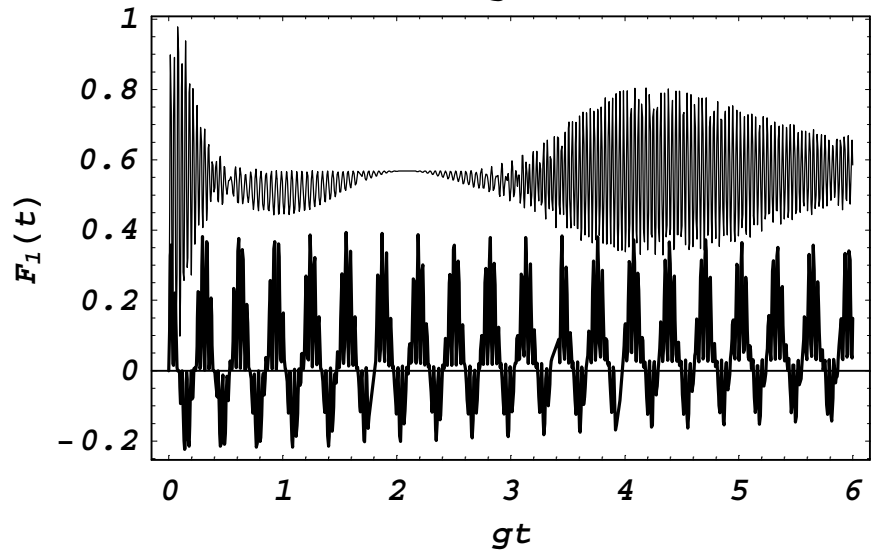

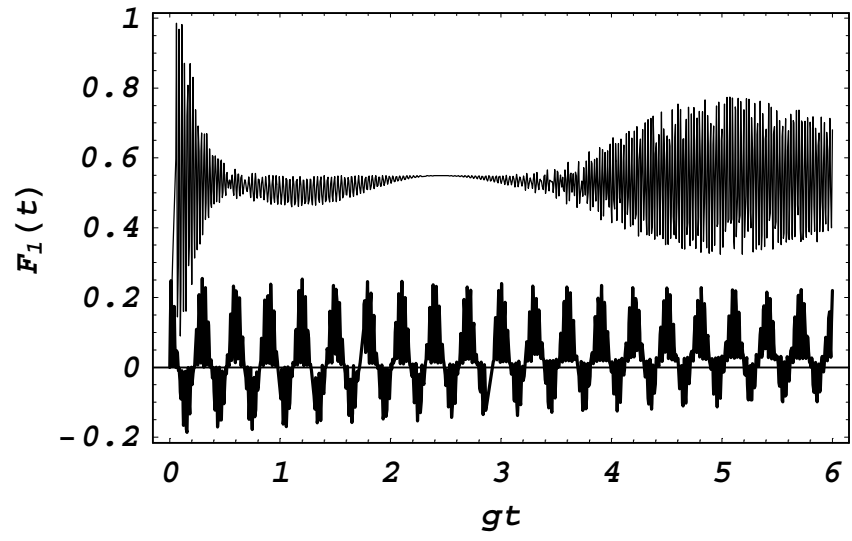

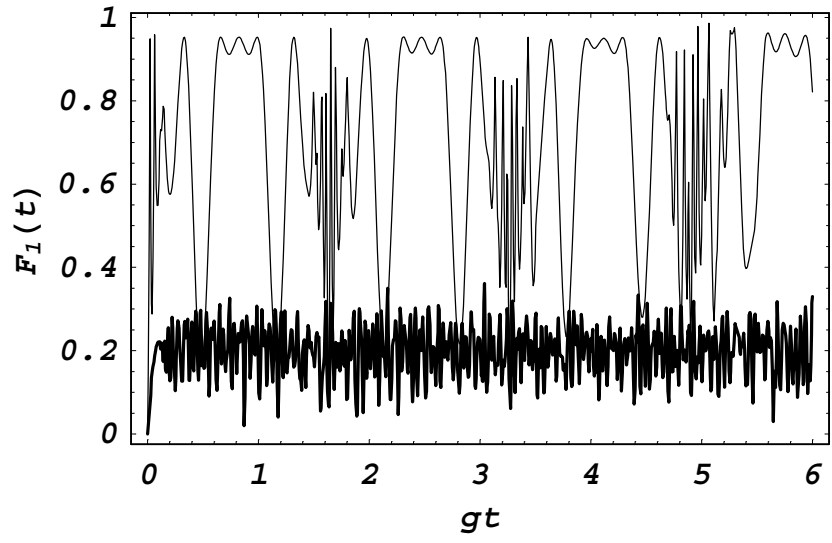